\documentclass[reprint,amsmath,amssymb,aps,superscriptaddress,twocolumn,10pt]{revtex4-1}
\usepackage[colorlinks=true,allcolors=blue]{hyperref}
\usepackage{graphicx,color}
\usepackage{dcolumn}
\usepackage{subcaption}
\usepackage{bm}
\usepackage{amsmath,amsfonts,amssymb}
\usepackage{acronym}
\usepackage{enumitem}
\usepackage{array}
\usepackage{multirow}
\usepackage{caption}
\allowdisplaybreaks
\newcommand{\be}{\begin{equation}}
\newcommand{\ee}{\end{equation}}
\newcommand{\bea}{\begin{eqnarray}}
\newcommand{\eea}{\end{eqnarray}}

\newcommand{\BNU}{School of Physics and Astronomy, Beijing Normal University, Beijing 100875, China.}
\newacro{GW}{gravitational wave}
\newacro{MBHB}{massive black hole binary}
\newacro{BH}{black hole}
\newacro{SIS}{singular isothermal sphere}
\newacro{NFW}{Navarro-Frenk-White}
\newacro{PE}{parameter estimation}
\newacro{SNR}{signal-to-noise ratio}
\newacro{PN}{post newtonion}
\newacro{FIM}{Fisher Information Matrix}
\newacro{GWTC}{Gravitational-wave Transient Catalog}
\begin{document}

\title{Effects of formation channels and gravitational lensing on stochastic gravitational wave background}

\author{Xin-yi Lin}
\email{xinyilin@bnu.edu.cn}
\affiliation{\BNU}

\author{Zhengxiang Li}
\email{zxli918@bnu.edu.cn}
\affiliation{\BNU}

\date{\today}

\begin{abstract}

Two primary formation channels for black holes have been proposed: the astrophysical channel, driven by the collapse of massive stars, and the primordial channel, involving their direct formation from density fluctuations in the early Universe. The key distinction between astrophysical black holes (ABHs) and primordial black holes (PBHs) is that PBHs can form at very high redshifts, before any stars have formed, leading to different stochastic gravitational-wave backgrounds (SGWBs). These SGWBs arise from the superposition of unresolved gravitational-wave signals accumulated over all redshifts. In this work, we employ the Hierarchical Bayesian Inference (HBI) framework and the publicly available GWTC-4 data to infer the population hyperparameters of PBHs. We then compute the SGWBs from ABHs and PBHs separately, accounting for the lensing effect, which can modify the strain amplitude of the SGWBs. By comparing the resulting SGWBs with the power-law integrated (PI) sensitivity curves of ground-based gravitational-wave detectors---LIGO and the Einstein Telescope (ET)---we find that both detectors can distinguish between these two black hole formation models within specific frequency ranges. However, LIGO is limited to a single method for distinguishing these models, and the lensing effect alters the frequency range over which discrimination is possible. In contrast, ET is capable of distinguishing ABHs from PBHs across a broader parameter space.

\end{abstract}

\maketitle

\section{Introduction}

Black holes may originate from two distinct formation channels: an astrophysical channel associated with stellar collapse, and a primordial channel in which they form in the early Universe from primordial fluctuations~\citep{Zeldovich:1967lct,Hawking:1971ei,1975ApJ...201....1C,Khlopov:1980mg,Khlopov:1985fch,Carr:2005zd,Sasaki:2018dmp,Mukherjee:2021ags}. Primordial black holes (PBHs) are considered a potential dark matter candidate~\citep{Frampton:2010sw,10.1093/mnras/152.1.75,1975ApJ...201....1C,10.1093/mnras/168.2.399,Chapline:1975ojl,Kawasaki:1997ju,Kohri:2007qn,Khlopov:2008qy,Carr:2021bzv,Green:2020jor,PhysRevD.104.083515}. However, no observational evidence currently supports the existence of a PBH population in the Universe.

Merging stellar-mass binary black holes (BBHs) are among the primary sources of detectable gravitational waves (GWs). The redshift distribution of the GW population, together with its correlations with source parameters, provides important constraints on the formation channels through which these binaries evolve~\citep{Fishbach:2017zga,Roulet:2018jbe,Gerosa:2017kvu,Kimball:2020opk,Talbot:2017yur,Fishbach:2018edt,Gerosa:2018wbw,Baibhav:2020xdf}. Notably, this population consists not only of individually detectable BBH mergers, but also of numerous distant events that remain unresolved~\citep{Christensen:2018iqi}. These unresolved signals cannot be detected separately, but they accumulate over all redshifts, forming a stochastic gravitational-wave background (SGWB)~\citep{Romano2017,Zhu:2011bd,Liang:2024ulf,Liang:2024tgn}. Thus, the SGWB can be produced by the superposition of unresolved populations of either primordial~\citep{Regimbau:2011rp} or astrophysical~\citep{1975JETP...40..409G} origin. The key distinction between astrophysical black holes (ABHs) and PBHs that enables their discrimination via the SGWB is that PBHs can form at very high redshifts, before any stars have formed, leading to different SGWB signatures.

As a GW propagates through a gravitational field, gravitational lensing modifies its strain amplitude~\citep{Ohanian:1974ys,Schneider:2006,Nakamura:1997sw,Nakamura:1999uwi,Takahashi:2003ix,Liao:2022gde,Oguri:2019fix,Leung:2023lmq} by a multiplicative magnification factor $\sqrt{\mu}$. This lensing-induced magnification can lead to biased inferences of the luminosity distance and chirp mass of GW sources~\citep{Dai:2017huk,Treu:2010uj}. Consequently, the SGWB inferred from observations of lensed events may differ from the true underlying background~\citep{Buscicchio:2020cij}.

Since the first GW signal~\citep{LIGOScientific:2016aoc} was detected by Advanced LIGO~\citep{LIGOScientific:2014pky,LIGOScientific:2016emj} and Virgo~\citep{VIRGO:2014yos} on September 14, 2015, the LIGO-Virgo-KAGRA (LVK) Collaboration has detected more than 200 GW signals by the time of GWTC-4~\citep{LIGOScientific:2025snk,LIGOScientific:2025hdt,LIGOScientific:2025yae,LIGOScientific:2025slb,LIGOScientific:2025pvj}. This rapidly growing catalog provides an unprecedented opportunity to investigate the properties, formation channels, and cosmological applications of GW sources. Using data from Advanced LIGO and Advanced Virgo spanning O1 through the first part of O4, the upper limit on the SGWB has been constrained to $\Omega_{\rm GW}(25\,{\rm Hz}) \le 2.0 \times 10^{-9}$~\citep{LIGOScientific:2025bgj}. Although no statistically significant SGWB signal has been detected by the LVK Collaboration to date, the steady improvement in detector sensitivity suggests that the SGWB may be observed in future observing runs. In addition, third-generation observatories such as the Einstein Telescope (ET)~\citep{Branchesi:2023mws,Punturo:2010zza}, with their lower detection thresholds, are expected to become operational in the near future, thereby significantly increasing the prospects for detecting the SGWB.

In this work, we first employ the hierarchical Bayesian inference (HBI) framework to infer the hyperparameters of the PBH population using the latest 142 BBH events from O1 through the first part of O4. We then aim to distinguish ABHs from PBHs through the SGWB using ground-based GW detectors LIGO and ET, while also accounting for the lensing effect on the SGWB.

This paper is organized as follows. In Sec.~\ref{Lensing effect on the SGWB}, we present the formalism of the SGWB and discuss the lensing effect. In Sec.~\ref{Merger rate density distribution of ABH binaries}, we introduce the merger-rate density distribution of ABH binaries. In Sec.~\ref{Constraints on the PBH population}, we present the method for constraining the parameters of PBHs. In Sec.~\ref{Merger rate density distribution of PBH binaries}, we describe the merger-rate density distribution of PBH binaries. In Sec.~\ref{Result}, we compare the lensed and unlensed SGWBs from ABHs and PBHs. Finally, in Sec.~\ref{Conclusions}, we summarize our conclusions and discuss the related implications.

\section{SGWB and the Lensing fraction}
\label{Lensing effect on the SGWB}
The energy–density spectrum of a GW background can be characterized as\cite{Zhu:2011bd,Zhu:2012xw}:
\begin{equation}
\Omega_{\mathrm{GW}}(\nu) = \frac{\nu}{\rho_c} \frac{d\rho_{\mathrm{GW}}}{d\nu},
\end{equation}
where $d\rho_{\mathrm{GW}}$ is the energy density in the frequency interval $\nu$ to $\nu+d\nu$, $\rho_c = 3H_0^2 c^2 / (8\pi G)$ is the critical energy density of the universe, and $H_0$ is the Hubble constant. For the binary mergers, the SGWB can be further transformed to\cite{Chen:2018rzo}
\begin{equation}
\begin{split}
\label{Omega_gw}
\Omega_{\rm GW}(\nu)&=
\frac{\nu}{\rho_c H_0}\int_0^{z_{\max}} dz \int dm_1\,dm_2 \\
&\quad \times\frac{\mathcal{R}(z,m_1,m_2)\,\frac{dE_{\rm GW}}{d\nu_s}(\nu_s,m_1,m_2)}{(1+z)\,E(\Omega_r,\Omega_m,\Omega_\Lambda,z)}.
\end{split}
\end{equation}
where $\nu_s=(1+z)\nu$ is the frequency, and
\begin{equation}
E(\Omega_r,\Omega_m,\Omega_\Lambda,z)=\sqrt{\Omega_r(1+z)^4+\Omega_m(1+z)^3+\Omega_\Lambda},
\end{equation}
where we take $\Omega_r=9.15\times10^{-5}$\cite{Fixsen:1996nj}, $\Omega_m=0.3111$ and $\Omega_\Lambda=0.6889$\cite{Planck:2018vyg}. The inspiral–merger–ringdown spectrum for a BBH is given by
\begin{equation}
\begin{split}
\frac{dE_{\mathrm{GW}}}{d\nu_s} &= \frac{(\pi G)^{2/3} M^{5/3} \eta}{3}\\
&\times
\begin{cases}
\nu_s^{-1/3}, & \nu_s < \nu_1, \\
\frac{\nu_s}{\nu_1} \nu^{-1/3}, & \nu_1 \leqslant \nu_s < \nu_2, \\
\frac{\nu_s^2}{\nu_1 \nu_2^{4/3}} \frac{\nu_4^4}{(4(\nu_s - \nu_2)^2 + \nu_4^2)^2}, & \nu_2 \leqslant \nu_s < \nu_3,
\end{cases}
\end{split}
\end{equation}
where $\nu_i = (a_i \eta^2 + b_i \eta + c_i) / (\pi G M / c^3)$, $M=m_1+m_2$ is the total mass of the BBH system, and $\eta=m_1 m_2/M^2$ is the symmetric mass ratio. The coefficients $a_i$, $b_i$ and $c_i$ are given in Table 1 of Ajith et al\cite{Ajith:2007kx}.

When a GW signal is perturbed by a cosmological lens, the mass-redshift degeneracy is reintroduced by an unknown lensing magnification $\mu$. Since the strain amplitude scales by a multiplicative factor of $\sqrt{\mu}$, the apparent mass $\tilde{M}$, redshift $\tilde{z}$, and luminosity distance $\tilde{d}_L$ are related to their true values by the following relationships\cite{Buscicchio:2020cij}:
\begin{equation}
d_L(\tilde{z}) = \frac{d_L(z)}{\sqrt{\mu}}, \quad \tilde{M}(1+\tilde{z}) = M(1+z).
\end{equation}
To incorporate lensing effects into the SGWB analysis, we modify Eq.\ref{Omega_gw} by replacing $\{dz,dm_{1,2}\}\rightarrow \{d\tilde{z}, d\tilde{m}_{1,2}\}$. Consequently, the apparent differential merger rate becomes $d^3\tilde{R}$, and the apparent redshifted frequency is given by $\nu_r = \nu(1+\tilde{z})$.

In this work, we approximate the lenses as singular isothermal spheres (SIS). When a GW event is identified as lensed, the distribution of the impact parameter $y$ in the range $y \in [0, y_0]$ follows:
\begin{equation}
P(y) = \frac{\tau(m_1, z_1, y)}{\int_0^{y_0} dy \, \tau(m_1, z_1, y)} = \frac{2}{y_0^2} y.
\end{equation}
To ensure a significant lensing effect, we set $y_0=1$, which yields the probability distribution $P(y)=2y$. For an SIS lens with $y \in [0, 1]$, double images are formed, and the magnification $\mu$ is related to the impact parameter $y$ by $\mu_{\pm} = \pm 1 + 1/y$\cite{Takahashi:2003ix}. Consequently, the probability distribution of the magnification $\mu$ is derived as:
\begin{align*}
\text{For } \mu_+: & \quad P(\mu_+) = \frac{2}{(\mu - 1)^3}, \quad \mu \in [2, +\infty), \\
\text{For } \mu_-: & \quad P(\mu_-) = \frac{2}{(\mu - 1)^3}, \quad \mu \in (0, +\infty),
\end{align*}
where $\mu_+$ and $\mu_-$ denote the magnifications of the two images, respectively. As a consequence, the lensed SGWB can be written as
\begin{equation}
\Omega_{\rm L}(\nu)=\int_{\mu_-}^{\mu_+} \tilde{\Omega}_{\rm GW}(\nu_r) d\mu.
\end{equation}

\section{Merger rate density distribution of ABH binaries}
\label{Merger rate density distribution of ABH binaries}
The merger rate density of the stellar-origin BBHs is given by\cite{Chen:2018rzo}
\begin{equation}
\label{R_abh}
\mathcal{R}=N \int_{t_{\min}}^{t_{\max}}
\frac{R_{\rm birth}\!\left(t(z)-t_d,\,m_1\right)\times P_d(t_d)}{\min\!\left(m_1,\,m_{\max}-m_1\right)-m_{\min}}\,dt_d
\end{equation}
where $N$ is a normalization constant, $t(z)$ is the age of the universe at merger, $P_d\propto t_d^{-1}$ is the distribution of the delay time $t_d$, with $t_{\rm{min}}<t_d<t_{\rm{max}}$. The minimum delay time of a massive binary system to evolve until coalescence are set to $t_{\rm{min}}=50\rm{Myr}$, and the maximum delay time $t_{\rm{max}}$ is set to the Hubble time.

The most complicated part in Eq.\ref{R_abh} is the birth rate of sBHs, which is given by\cite{Dvorkin:2016wac}
\begin{equation}
\begin{split}
R_{\mathrm{birth}}(t, m_{\mathrm{rem}}) &= \int \psi[t - \tau(m_*)] \phi(m_*)\\
&\times \delta(m_* - g_{\mathrm{rem}}^{-1}(m_{\mathrm{rem}})) dm_*,
\end{split}
\end{equation}
where $m_*$ is the mass of the progenitor star, $m_{rem}$ is the mass of remnant, and $\tau(m_*)$ is the lifetime of a progenitor star, which is negligible\cite{Schaerer:2001jc}. And $\phi(m_*) \propto m_*^{-2.35}$ is the initial mass function, $\psi(t)$ is the star formation rate (SFR), which is given by\cite{Nagamine:2003bd}
\begin{equation}
\psi(z) = k \frac{a \exp[b(z - z_m)]}{a - b + b \exp[a(z - z_m)]}.
\end{equation}
In this work, we use the “\textit{Fiducial+PopIII}” model which is the sum of \textit{Fiducial} SFR (with $k = 0.178 \, M_\odot \, \mathrm{yr}^{-1} \, \mathrm{Mpc}^{-3}, \quad z_m = 2, \quad a = 2.37, b=1.8$) and \textit{PopIII} SFR (with $k = 0.002 \, M_\odot \, \mathrm{yr}^{-1} \, \mathrm{Mpc}^{-3}, \quad z_m = 11.87, \quad a = 13.8, b=13.36$). Additionally, we consider the \textit{WWp} model~\cite{Woosley:1995ip} of sBH formation. In this framework, the relationship of the progenitor with an initial mass $m_*$ and the mass of the remnant BH $m_{\mathrm{bh}}$ is given by
\begin{equation}
\frac{m_{\mathrm{bh}}}{m_*} = A \left( \frac{m_*}{40 M_\odot} \right)^\beta \frac{1}{\left( \frac{Z(z)}{0.01 Z_\odot} \right)^\gamma + 1},
\end{equation}
where $Z(z)$ is the metallicity. Following Dvorkin et al\cite{Dvorkin:2016wac}, we adopt the parameter values $A=0.3$, $\beta=0.8$ and $\gamma=0.2$.

\section{Merger rate density distribution of PBH binaries}
\label{Merger rate density distribution of PBH binaries}

In this section, we discuss three models of the merger rate density of PBHs in units of $\rm{Gpc}^{-3}\rm{yr}^{-1}$. Theoretically, two distinct mechanisms are proposed for the formation of PBH binaries. The first mechanism occurs via decoupling from cosmic expansion during the radiation-dominated era of the early Universe\cite{Sasaki:2016jop,Ali-Haimoud:2017rtz,Raidal:2017mfl,Chen:2018czv,Raidal:2018bbj,Chen:2024dxh,Raidal:2024bmm,Carr:2024nlv,Huang:2024wse,DeLuca:2025fln}. The second one is that the PBHs binary form in the late Universe by the close encounter\cite{Sasaki:2016jop,Raidal:2017mfl}. In this work, we apply the first formation mechanism because the first channel contribute the dominant GW sources of BBHs.

%We use the lognormal mass function.
In this work, we employ the lognormal PBH mass function for our analysis. This function serves as a versatile approximation for a wide range of extended mass distributions, particularly when PBHs originate from a smooth, symmetric peak in the inflationary power spectrum under the slow-roll approximation. A lognormal PBH mass function is defined by\cite{Dolgov:1992pu,Carr:2017jsz,Bellomo:2017zsr}
\begin{equation}
P(m) = \frac{1}{\sqrt{2\pi} \sigma m} \exp\left(-\frac{\ln^2(m/M_c)}{2\sigma^2}\right),
\end{equation}
where $M_c$ is the median mass and $\sigma$ characterizes the width of the mass distribution. A PBH population with a lognormal mass function requires three parameters to model the binary merger rate, $\Lambda=\{M_c,\sigma,f_{\rm{PBH}}\}$.

\subsection{The merger rate density of a PBH binary}
The merger rate density of a PBH binary is expressed as\cite{Chen:2018czv,Raidal:2018bbj,Liu:2018ess,Chen:2024dxh,Miller:2024fpo}
\begin{equation}
\label{Eq-null}
\begin{split}
\mathcal{R}(t,m_1,m_2)&=1.6\times10^6\left(\frac{t}{t_0}\right)^{-\frac{34}{37}}f_{\rm PBH}^{\frac{53}{37}}\eta^{-\frac{34}{37}}\\
&\times\left(\frac{M}{M_\odot}\right)^{-\frac{32}{37}} P(m_1)P(m_2)
\end{split}
\end{equation}
where $f_{\rm PBH}$ is the abundance of PBH in the dark matter, $t_0$ is the present cosmic time. The cosmic time $t$ and the redshift $z$ are related by
\begin{equation}
t(z) = \int_{z}^{\infty} \frac{dz'}{H(z')(1+z')}
\end{equation}
where $H(z)$ is the Hubble parameter.

\subsection{The merger rate density of a PBH binary with density perturbations}
Here we consider the merger rate density that taking into account the linear density perturbations\cite{Chen:2018czv,Nitz:2020bdb,Chen:2021nxo,Liu:2022elb},
\begin{equation}
\label{Eq-sigmaeq}
\begin{split}
\mathcal{R}(t,m_1,m_2)&=2.8\times10^{6}\left(\frac{t}{t_0}\right)^{-\frac{34}{37}}f_{\rm{PBH}}^{2}\left(0.7 f_{\rm{PBH}}^{2}+\sigma_{\rm eq}^{2}\right)^{-\frac{21}{74}} \\
& \times\min\!\left(\frac{P(m_1)}{m_1},\frac{P(m_2)}{m_2}\right)\times\left(\frac{P(m_1)}{m_1}+\frac{P(m_2)}{m_2}\right) \\
& \times(m_1m_2)^{\frac{3}{37}}(m_1+m_2)^{\frac{36}{37}} ,
\end{split}
\end{equation}
where $\sigma_{\rm eq}\approx0.005$ is the variance of density perturbations of the rest CDM on scale of order $\mathcal{O}(10^0 \sim 10^3) M_\odot$ at radiation-matter equality.

\subsection{The merger rate density of a PBH binary with a suppression factor}
Then we consider the merger rate density with a suppression factor\cite{Raidal:2018bbj,Zheng:2022wqo,Dizon:2024iao,Dizon:2025siw},
\begin{equation}
\label{Eq-s}
\begin{split}
\mathcal{R}(t,m_1,m_2,z)&=1.6\times10^6\left(\frac{t}{t_0}\right)^{-\frac{34}{37}}f_{\rm PBH}^{\frac{53}{37}}\eta^{-\frac{34}{37}}\\
&\times\left(\frac{M}{M_\odot}\right)^{-\frac{32}{37}} P(m_1)P(m_2)\\
&\times S\!\left(M,f_{\rm PBH},P(m),z\right).
\end{split}
\end{equation}
Here $S\!\left(M,f_{\rm PBH},P(m),z\right)$ denotes a suppression factor $S<1$ that accounts for two effects: the influence of the surrounding smooth matter component on PBH binary formation and the disruption of PBH binaries by nearby PBH clusters. We can define each contribution individually as
\begin{equation}
\label{S}
S \equiv S_1(M, f_{\rm PBH}, P_{\rm PBH}(m|\theta)) S_2(f_{\rm PBH}, z).
\end{equation}
where the first term $S_1$ is approximate to
\begin{equation}
\label{S1}
S_1 \approx 1.42 \left[ \frac{\langle m^2 \rangle / \langle m \rangle^2}{\bar{N} + C} + \frac{\sigma_{\rm eq}^2}{f_{\rm PBH}^2} \right]^{-\frac{21}{74}} \exp(-\bar{N}),
\end{equation}
with
\begin{equation}
\bar{N} \equiv \frac{M}{\langle m \rangle} \left( \frac{f_{\rm PBH}}{f_{\rm PBH} + \sigma_{\rm eq}} \right),
\end{equation}
where the constant factor $C$ is defined as
\begin{equation}
C \equiv \frac{\langle m^2 \rangle f_{\rm PBH}^2}{\langle m \rangle^2 \sigma_{\rm m}^2}\times\left\{ \left[ \frac{\Gamma(29/37)}{\sqrt{\pi}} U\left( \frac{21}{74}, \frac{1}{2}, \frac{5f_{\rm PBH}^2}{6\sigma_{\rm eq}^2} \right) \right] - 1 \right\}^{-1},
\end{equation}
where $\Gamma(x)$ and $U(a,b,z)$ Gamma function and confluent hypergeometric function respectively. And the mass average in Eq.\ref{S1} is defined as
\begin{equation}
\langle m^n \rangle \equiv \int m^n P_{\mathrm{PBH}}(m|\theta) dm.
\end{equation}
Accounting for the fraction of PBH binaries disrupted by other PBH clusters, the second term $S_2$ in Eq.\ref{S} is given by
\begin{equation}
S_2 \approx \min[1, 9.6 \times 10^{-3} x^{-0.65} \exp(0.03 \ln^2 x)],
\end{equation}
where $x \equiv (t(z)/t_0)^{0.44} f_{\rm{PBH}}$.

\section{Constraints on the PBH population}
\label{Constraints on the PBH population}

\begin{figure*}[t!]
    \centering
    \begin{subfigure}[b]{0.48\textwidth}
        \centering
        \includegraphics[width=\textwidth]{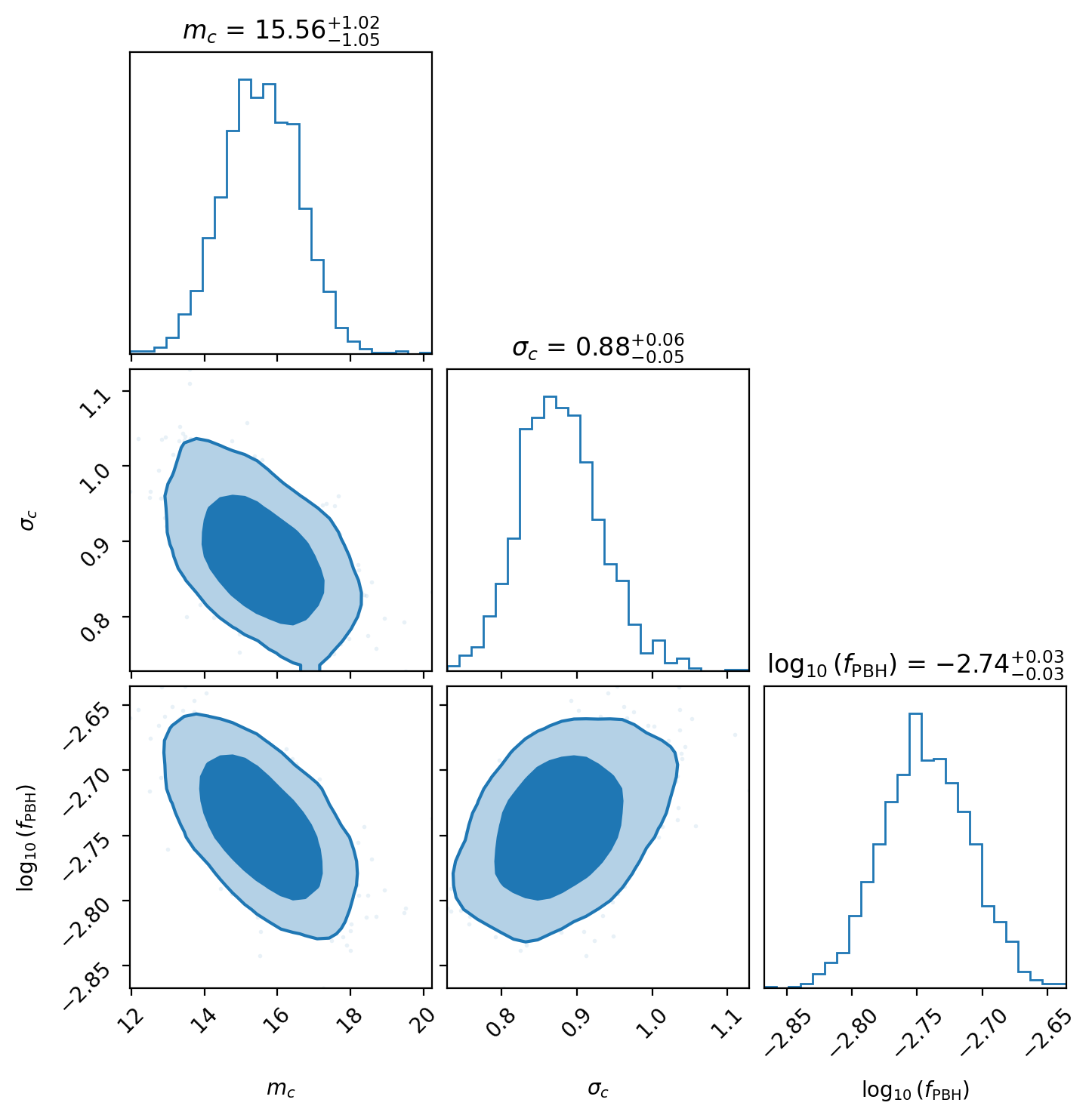}
        \caption{}
        \label{pbh_corner:null}
    \end{subfigure}
    \hfill
    \begin{subfigure}[b]{0.48\textwidth}
        \centering
        \includegraphics[width=\textwidth]{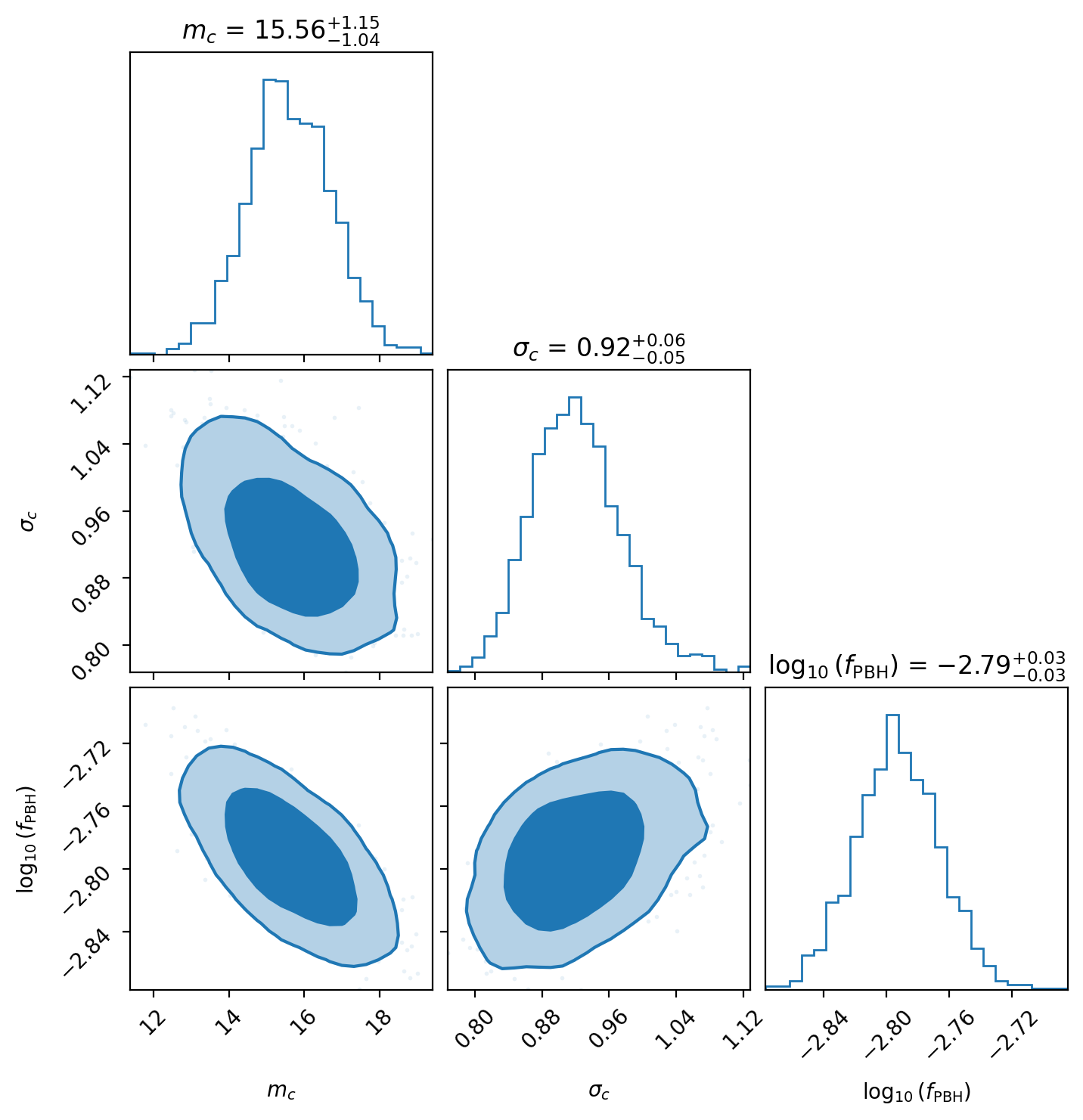}
        \caption{}
        \label{pbh_corner:sigmaeq}
    \end{subfigure}
    
    \leavevmode\\
    \vspace{0.1cm}
    \begin{subfigure}[b]{0.48\textwidth}
        \centering
        \includegraphics[width=\textwidth]{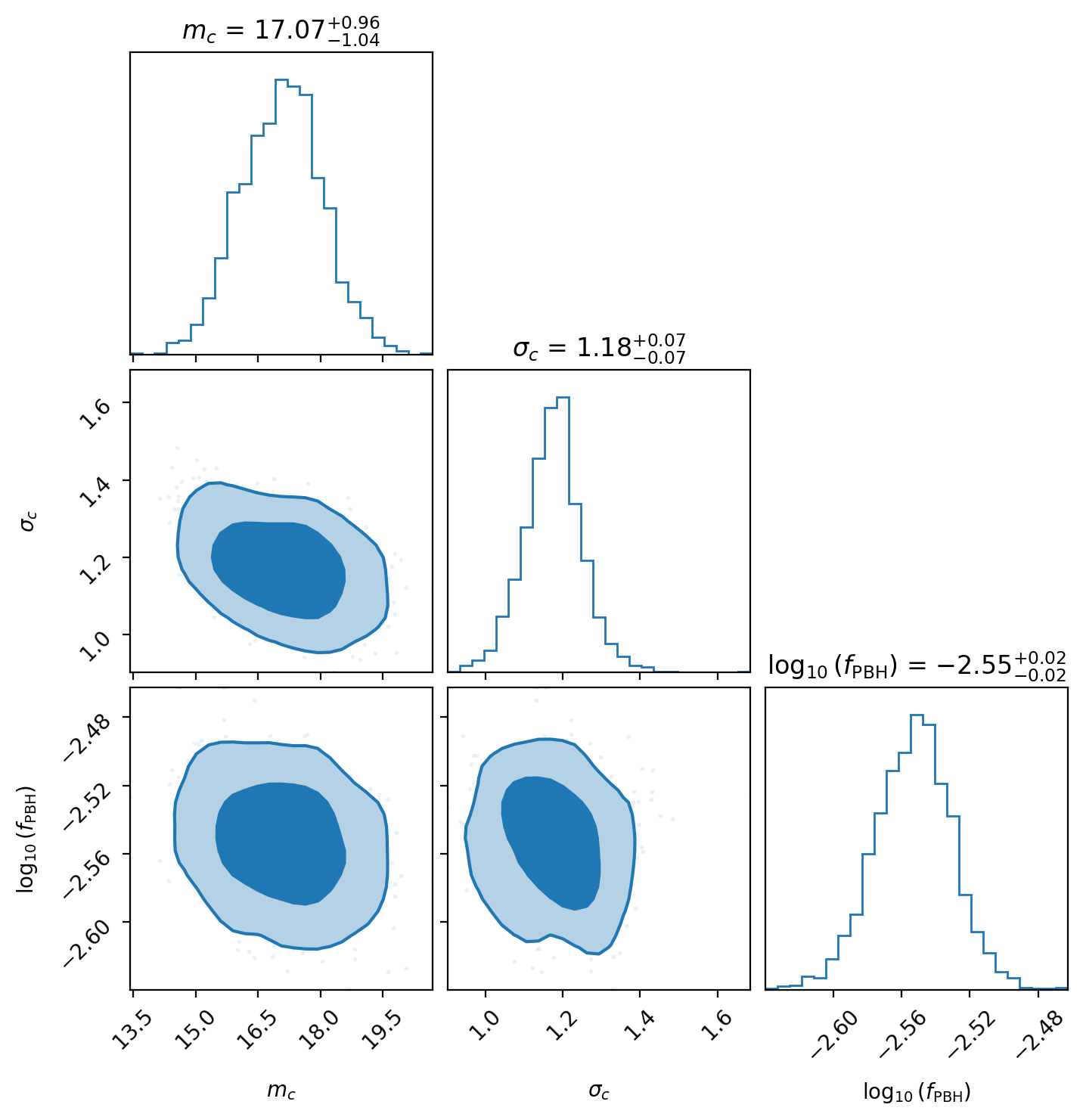}
        \caption{}
        \label{pbh_corner:s}
    \end{subfigure}
    \captionsetup{justification=raggedright, singlelinecheck=false}
    \caption{The posterior distributions of the hyperparameters for the different PBH models are presented. The upper left panel displays the posterior distributions derived from the PBH merger rate defined in Eq.\ref{Eq-null}. The upper right panel shows the posterior distributions derived from the PBH merger rate with density perturbations, as defined in Eq.\ref{Eq-sigmaeq}. The lower panel presents the posterior distributions derived from the PBH merger rate with suppression, as defined in Eq.\ref{Eq-s}.}
    \label{fig:double_col_layout}
\end{figure*}

In this work, we employ the HBI framework and the publicly available GWTC-4 data to determine the mass distribution function $P(m|\sigma_c; m_c)$ of the PBH model and the abundance of PBHs, $f_{\text{PBH}}$. The HBI method is commonly employed to infer the parameters of an underlying distribution from a set of observations subject to measurement uncertainties and selection effects. To infer the population hyperparameters from $N_{\text{obs}}$ GW evens $\mathbf{d} = \{d_1, \dots, d_{N_{\text{obs}}}\}$, the likelihood for the observed BBH events can be expressed as
\begin{equation}
\begin{split}
p(\mathbf{d}|\Phi) &\propto N(\Phi)^{N_{\text{obs}}} e^{-N(\Phi)\xi(\Phi)}\\
& \times\prod_{i}^{N_{\text{obs}}} \int d\lambda \mathcal{L}(d_i|\lambda) p_{\text{pop}}(\lambda|\Phi),
\end{split}
\end{equation}
where the likelihood of one BBH event $\mathcal{L}(d_i|\lambda)$ is proportional to the posterior $p(\lambda|d_i)$. $N(\Phi)$ is the total number of events in the model characterized by the set of population parameters $\Phi$ as
\begin{equation}
N(\Phi) = \int d\lambda T_{\text{obs}} \mathcal{R}(\lambda|\Phi) \frac{1}{1+z} \frac{dV_c}{dz},
\end{equation}
where $dV_c/dz$ is the differential comoving volume, the factor $1/(1+z)$ is the cosmological time dilation from the source frame to the detector frame, and $T_{\text{obs}}$ is the effective observing time of LIGO O1 to the first part of O4 running. $p_{\text{pop}}(\lambda|\Phi)$ is the normalized distribution of black hole masses and redshifts in coalescing binaries as
\begin{equation}
p_{\text{pop}}(\lambda|\Phi) = \frac{1}{N(\Phi)} \left[ T_{\text{obs}} \mathcal{R}(\lambda|\Phi) \frac{1}{1+z} \frac{dV_c}{dz} \right].
\end{equation}
Meanwhile, $\xi(\Phi)$ is defined as
\begin{equation}
\xi(\Phi) \equiv \int d\lambda P_{\text{det}}(\lambda) p_{\text{pop}}(\lambda|\Phi),
\end{equation}
where $P_{\text{det}}(\lambda)$ is the detection probability that depends on the source parameters $\lambda$.
$\xi(\Phi)$ is estimated using a Monte Carlo integral over the detected injections,
\begin{equation}
\xi(\Phi) \approx \frac{1}{N_{\text{inj}}} \sum_{k=1}^{N_{\text{det}}} \frac{p_{\text{pop}}(\lambda_k|\Phi)}{p_{\text{draw}}(\lambda_k)},
\end{equation}
where $N_{\text{inj}}$ denotes the total number of injections, $N_{\text{det}}$ represents the number of recovered injections, and $p_{\text{draw}}$ is the prior distribution from which the injection parameters are sampled. Then the posterior distribution $p(\Phi|d)$ can be calculated by
\begin{equation}
p(\Phi|d) = \frac{p(d|\Phi)p(\Phi)}{Z_{\mathcal{M}}},
\end{equation}
where $p(\Phi)$ is prior distribution for the population hyperparameters $\Phi$. The prior distributions for the
population hyperparameters are exhibited in Table \ref{tab:priors}.
\begin{table}[h]
\centering
\caption{Prior distributions for the population hyperparameters of PBHs.}
\label{tab:priors}
\begin{tabular}{|c|c|}
\hline
Hyperparameter $\Phi$ & Prior \\ \hline
$m_{\text{c}}$ & $\mathcal{U}[5, 50]$ \\ \hline
$\sigma_{\text{c}}$ & $\mathcal{U}[0.1, 2]$ \\ \hline
$f_{\text{PBH}}$ & $\log-\mathcal{U}[-4, 0]$ \\ \hline
\end{tabular}
\end{table}

Incorporating the three different PBH population models and 142 BBH events from GWTC-4 into the $\texttt{ICAROGW}$, we estimate the posterior distributions of the hyperparameters $\Phi = [m_c, \sigma_c, \mathrm{log}_{10}(f_{\mathrm{PBH}})]$. Fig.\ref{pbh_corner:null} and Fig.\ref{pbh_corner:s} present the posterior distributions of the hyperparameters for the PBH models defined by Eqs.\ref{Eq-null} and Eqs.\ref{Eq-s}, respectively. For the first model, our analysis yields the following median estimates with $68\%$ equal-tailed credible intervals: $m_c=15.56^{+1.02}_{-1.05}$, $\sigma_c=0.88^{+0.06}_{-0.05}$, and $\mathrm{log}_{10}(f_{\mathrm{PBH}})=-2.74^{+0.03}_{-0.03}$. These results are comparable to the posterior distributions obtained from GWTC-3 data in Chen et al.\cite{Chen:2024dxh}. Similarly, for the second model, we find $m_c=15.56^{+1.15}_{-1.04}$, $\sigma_c=0.92^{+0.06}_{-0.05}$, and $\mathrm{log}_{10}(f_{\mathrm{PBH}})=-2.79^{+0.03}_{-0.03}$, which aligns well with the results derived from GWTC-3 by Zheng et al.\cite{Zheng:2022wqo}. Fig.\ref{pbh_corner:sigmaeq} displays the posterior distributions of the hyperparameters for the PBHs models defined by Eq.\ref{Eq-sigmaeq}. Our analysis yields hyperparameter values with median estimates and $68\%$ equal-tailed credible intervals: $m_c=17.07^{+0.96}_{-1.04}$, $\sigma_c=1.18^{+0.07}_{-0.07}$, and $\mathrm{log}_{10}(f_{\mathrm{PBH}})=-2.55^{+0.02}_{-0.02}$.

\section{Result}
\label{Result}

In this work, we compare the SGWB generated by different black holes with the PI curves for ET-D with 4 years of observation and LIGO’s observing runs of O4. For ABHs, we focus on stellar-mass black holes with masses in the range of $5\sim100\,M_{\odot}$ and redshifts of $0\sim20$. For PBHs, we adopt the same mass range $5\sim100\,M_{\odot}$ but extend the redshift range to $0\sim1000$, accounting for their existence at very high redshifts. Furthermore, we account for the lensing effect on the SGWB. We restrict the integration ranges to $\mu_+ \in [2, 20]$ and $\mu_- \in [0, 20]$. We have verified that the probability of $\mu_{\pm} > 20$ is of the order $\mathcal{O}(10^{-3})$, which is small enough to be neglected.

\subsection{The SGWB of ABHs and PBHs}
\label{result_null}

\begin{figure}[h]
\centering
\includegraphics[width=0.5\textwidth]{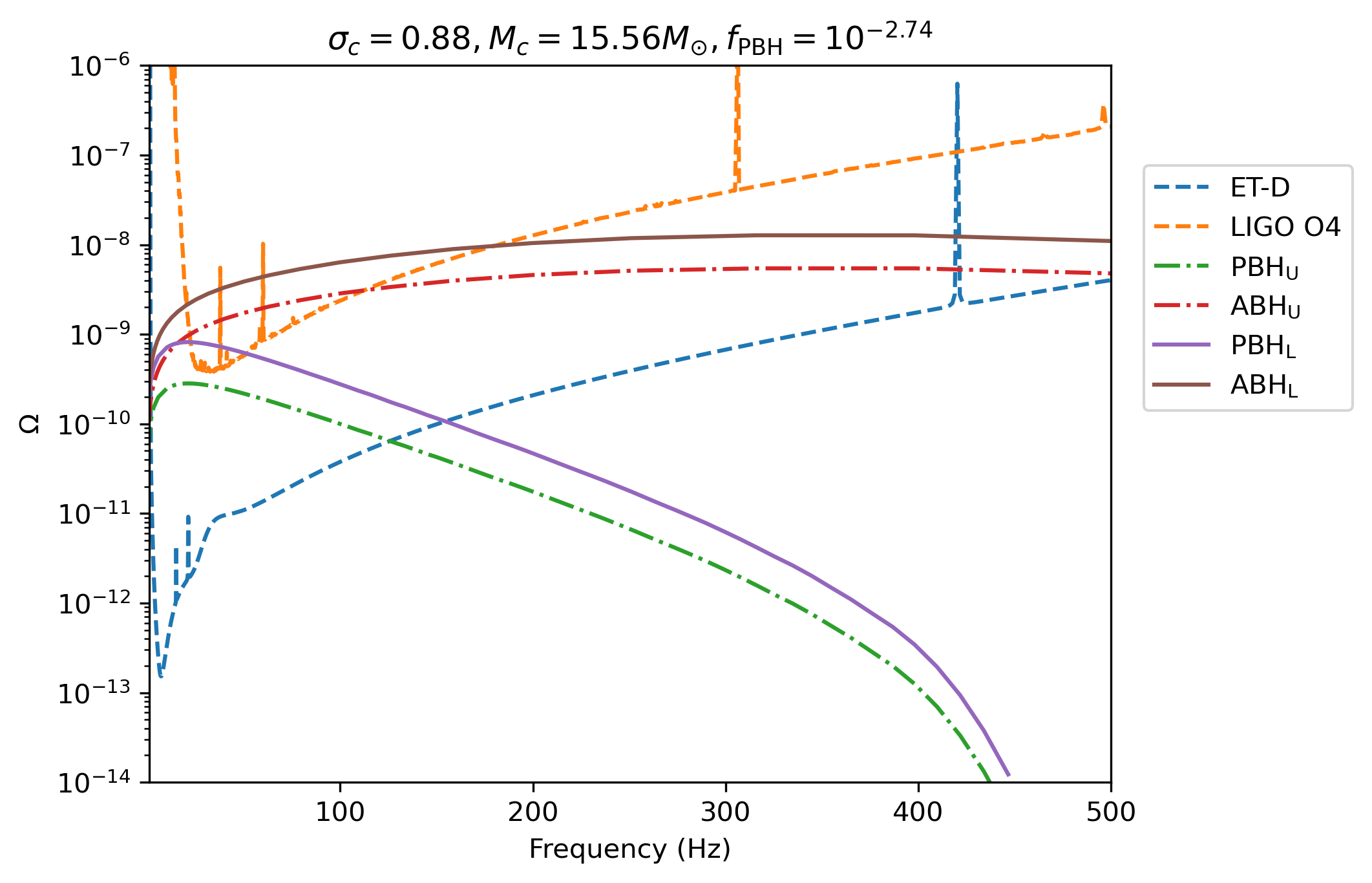}
\captionsetup{justification=raggedright, singlelinecheck=false}
\caption{The SGWB generated by ABHs and PBHs under the PBHs merger rate model given in Eq.\ref{Eq-null}. The orange and the blue dash curves are the sensitivity curve of LIGO O4 and ET-D, respectively. The red and the green dash-dot lines represent the unlensed SGWB generated by ABHs and PBHs respectively. And the brown and the purple solid lines represent the lensed SGWB generated by ABHs and PBHs respectively.} 
\label{null_ground}
\end{figure}

We displays the SGWB generated by ABHs and PBHs under the PBHs merger rate model given in Eq.\ref{Eq-null} in Fig.\ref{null_ground}. For this analysis, we adopt the parameters $m_c=15.56$, $\sigma_c=0.88$, and $\mathrm{log}_{10}(f_{\mathrm{PBH}})=-2.74$. In the frequency range of $22\sim113\,\rm{Hz}$, the unlensed SGWB for ABHs exceeds the sensitivity curves of LIGO O4, indicating that it is detectable by LIGO. In contrast, the unlensed SGWB for PBHs curve lies entirely below the PI curves, implying that the SGWB from unlensed PBHs remains undetectable. Consequently, any SGWB detected in this case would only originate from ABHs. Therefore, in the unlensed scenario, LIGO can effectively distinguish ABHs from PBHs via the SGWB in the $22\sim113\,\rm{Hz}$ frequency range.

In the lensed scenario, within the frequency range of $20\sim180\,\rm{Hz}$, the lensed SGWB from ABHs exceeds the LIGO O4 sensitivity curve, while the lensed SGWB from PBHs also surpasses these curves in the $22\sim51\,\rm{Hz}$ band. This implies that LIGO cannot distinguish between these two black hole populations in the overlapping frequency range. However, in the $51\sim180\,\rm{Hz}$ range, the lensed SGWB from ABHs remains above the LIGO O4 sensitivity curves, whereas the signal from PBHs falls below them. This indicates that LIGO can effectively distinguish between the two populations in this frequency band. Compared to the unlensed case, the frequency range in which LIGO can discriminate between ABHs and PBHs becomes slightly broader.

We now focus on the sensitivity curves of ET-D. In the frequency range of $5\sim126\,\rm{Hz}$, the unlensed SGWBs from both ABHs and PBHs lie above the ET-D sensitivity curves, yet their amplitudes differ significantly. Beyond this range, the unlensed SGWB from ABHs remains above the ET-D curves, whereas the signal from PBHs falls below them. Consequently, ET can distinguish between ABHs and PBHs based on their signal strengths within the $5\sim126\,\rm{Hz}$ band, and can exclusively detect the unlensed SGWB from ABHs at frequencies above $126\,\rm{Hz}$.

For the lensed case, the SGWBs from both ABHs and PBHs lie above the ET-D sensitivity curves in the $5\sim155\,\rm{Hz}$ range with obviously different signal strengths, whereas only the SGWB from ABHs remains detectable beyond this band. We conclude that ET can distinguish between ABHs and PBHs based on their signal strengths within the $5\sim155\,\rm{Hz}$ band, and exclusively detect the SGWB from ABHs at frequencies above $155\,\rm{Hz}$. Compared to the unlensed case, the frequency range for discriminating between ABHs and PBHs by signal strengths becomes slightly broader, while the frequency range for only detecting the SGWB from ABHs becomes narrower.

\subsection{The SGWB of ABHs and PBHs binaries with density perturbations}
\label{result_sigmaeq}

\begin{figure}[h]
\centering
\includegraphics[width=0.5\textwidth]{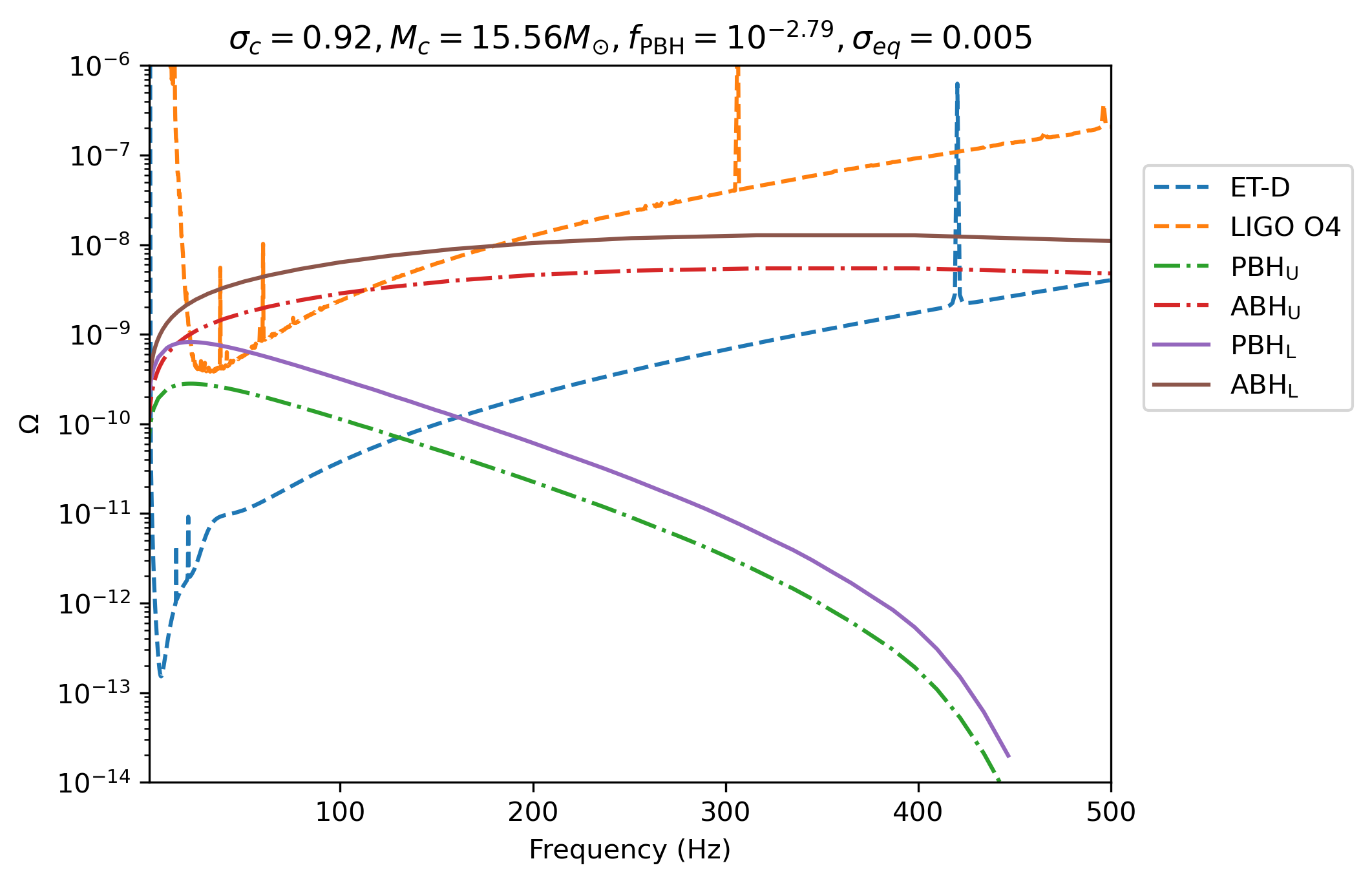}
\captionsetup{justification=raggedright, singlelinecheck=false}
\caption{The SGWB generated by ABHs and PBHs under the PBHs merger rate with density perturbations. The orange and the blue dash curves are the sensitivity curve of LIGO O4 and ET-D, respectively. The red and the green dash-dot lines represent the unlensed SGWB generated by ABHs and PBHs respectively. And the brown and the purple solid lines represent the lensed SGWB generated by ABHs and PBHs respectively.} 
\label{sigma_eq_ground}
\end{figure}

Fig.\ref{sigma_eq_ground} shows the SGWB generated by ABHs and PBHs under the PBHs merger rate that with density perturbations. For this analysis, we adopt the parameters $m_c=15.56$, $\sigma_c=0.92$, and $\mathrm{log}_{10}(f_{\mathrm{PBH}})=-2.79$ and the the variance of density perturbations $\sigma_{\rm{eq}}=0.005$.

We observe that in the frequency range of $22\sim113\,\rm{Hz}$, the unlensed SGWB curve for ABHs exceeds the sensitivity curves of LIGO O4, indicating that LIGO can distinguish ABHs from PBHs in this band in the unlensed scenario. This finding is consistent with the case in \ref{result_null}. In contrast, in the lensed scenario, the lensed SGWB of ABHs exceeds the LIGO O4 sensitivity curves in the $20\sim180\,\rm{Hz}$ range, whereas the lensed SGWB from PBHs surpasses these curves only in the $22\sim53\,\rm{Hz}$ band. This differs slightly from the case in \ref{result_null}.

For the sensitivity curve of ET-D, in the frequency range of $5\sim131\,\rm{Hz}$, the unlensed SGWBs from both ABHs and PBHs lie above the ET-D sensitivity curve while their amplitudes are significantly different. Beyond this range, only the unlensed SGWB from ABHs remains above the ET-D curves. Resultly, ET can distinguish between ABHs and PBHs based on the strengths of SGWB within the $5\sim131\,\rm{Hz}$ band, and can exclusively detect the unlensed SGWB from ABHs at frequencies above $131\,\rm{Hz}$. For lensed scenario, the SGWBs from both ABHs and PBHs lie above the ET-D sensitivity curves in the $5\sim161\,\rm{Hz}$ range with obviously different signal strengths, whereas only the SGWB from ABHs remains detectable beyond this band. So ET can distinguish between ABHs and PBHs based on the strengths of lensed SGWBs within the $5\sim161\,\rm{Hz}$ band, and can exclusively detect the lensed SGWB from ABHs beyond $161\,\rm{Hz}$. The frequency range for only detecting the SGWB from ABHs becomes narrower and the frequency range for distinguishing these two black holes by the strengths of SGWBs becomes wider. These results are slightly different from the case in \ref{result_null}, too.

\subsection{The SGWB of ABHs and PBHs binaries with a  suppression factor}
\begin{figure}[h]
\centering
\includegraphics[width=0.5\textwidth]{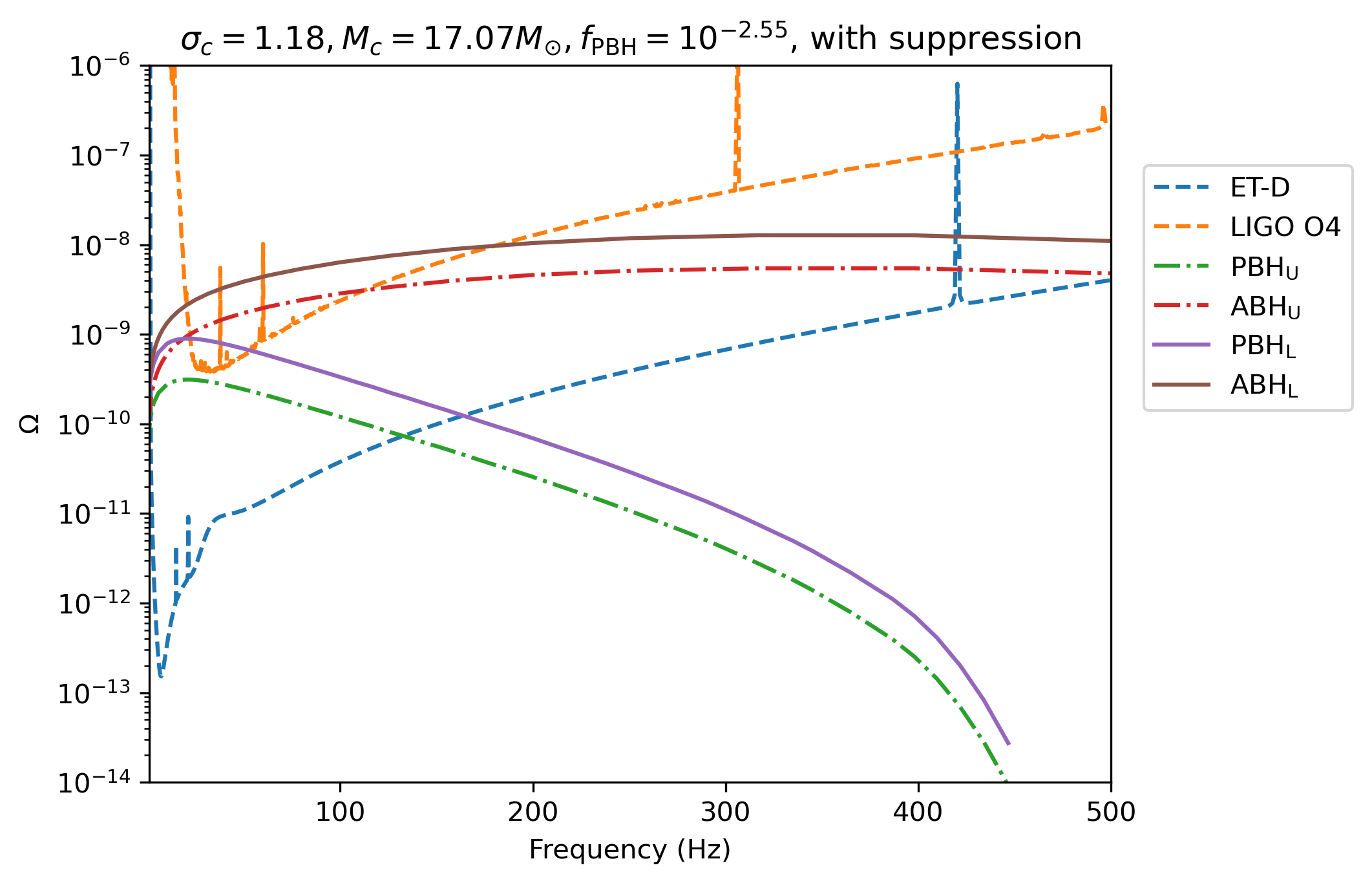}
\captionsetup{justification=raggedright, singlelinecheck=false}
\caption{The SGWB generated by ABHs and PBHs under the PBHs merger rate with a supression factor. The orange and the blue dash curves are the sensitivity curve of LIGO O4 and ET-D, respectively. The red and the green dash-dot lines represent the unlensed SGWB generated by ABHs and PBHs respectively. And the brown and the purple solid lines represent the lensed SGWB generated by ABHs and PBHs respectively.} 
\label{suppression_ground}
\end{figure}

Fig.\ref{suppression_ground} shows the SGWB generated by ABHs and PBHs under the PBHs merger rate that with a suppression. For this analysis, we adopt the parameters $m_c=17.07$, $\sigma_c=1.18$, and $\mathrm{log}_{10}(f_{\mathrm{PBH}})=-2.55$ and the rescaled variance of matter density perturbations $\sigma_{\rm{eq}}=0.005$.

The same as the case presented in \ref{result_null} and \ref{result_sigmaeq}, in the unlensed scenario, LIGO is capable of detecting the SGWB from ABHs within the frequency range of $22\sim113\,\rm{Hz}$. However, in the lensed scenario, the frequency range in which LIGO can distinguish ABHs from PBHs is $53\sim180\,\rm{Hz}$, which differs slightly from the results we presented in \ref{result_null}.

For the ET-D curve, in the unlensed scenario, the ET detector can distinguish between ABHs and PBHs based on the strength of the SGWB in the frequency range of $5\sim133\,\rm{Hz}$; beyond this range, only the SGWB from ABHs is detectable. In the lensed scenario, ET can distinguish these two types of black holes within the frequency range of $5\sim164\,\rm{Hz}$, while only the SGWB from ABHs can be detected beyond this limit.

The results for this model are similar to those for the two models discussed previously. For the LIGO detector, the lensing effect expands the frequency range to some extent, whereas for ET, the lensing effect alters the frequency boundaries of the different judgment criteria.

\section{Conclusions}
\label{Conclusions}

In this paper, we propose that ground-based GW detectors can distinguish between ABHs and PBHs via the SGWB, and we incorporate the lensing effect into our analysis. We adopt the ``Fiducial+PopIII'' model for the ABH merger rate density and a lognormal mass function for PBHs. For the PBH merger rate density, we consider three cases: (i) without any additional effects; (ii) including density perturbations; and (iii) further incorporating a suppression factor. Using the HBI framework and the publicly available GWTC-4 data, we estimate the posterior distributions of the hyperparameters $\Phi = [m_c, \sigma_c, \log_{10}(f_{\mathrm{PBH}})]$ with $\texttt{ICAROGW}$. For the three models, the median estimates and $68\%$ equal-tailed credible intervals are:

\begin{itemize}
    \item Model~(i): $m_c = 15.56^{+1.02}_{-1.05}$, $\sigma_c = 0.88^{+0.06}_{-0.05}$, $\log_{10}(f_{\mathrm{PBH}}) = -2.74^{+0.03}_{-0.03}$;
    \item Model~(ii): $m_c = 15.56^{+1.15}_{-1.04}$, $\sigma_c = 0.92^{+0.06}_{-0.05}$, $\log_{10}(f_{\mathrm{PBH}}) = -2.79^{+0.03}_{-0.03}$;
    \item Model~(iii): $m_c = 17.07^{+0.96}_{-1.04}$, $\sigma_c = 1.18^{+0.07}_{-0.07}$, $\log_{10}(f_{\mathrm{PBH}}) = -2.55^{+0.02}_{-0.02}$.
\end{itemize}

Using these values and accounting for the lensing fraction, we compute the SGWBs for different black hole populations, both with and without lensing.

We compare the SGWBs from different black hole models with the PI curves for ET-D (4-year observation) and LIGO's O4 run. In the unlensed scenario, only the ABH-generated SGWB is detectable within $22{-}113\,\mathrm{Hz}$ for all PBH models. In the lensed scenario, LIGO can distinguish ABHs from PBHs in the following frequency ranges: $51{-}180\,\mathrm{Hz}$ for Model~(i), and $53{-}180\,\mathrm{Hz}$ for both Models~(ii) and~(iii).

For ET-D in the unlensed scenario, the distinguishable frequency ranges are $5{-}126\,\mathrm{Hz}$ (Model~i), $5{-}131\,\mathrm{Hz}$ (Model~ii), and $5{-}133\,\mathrm{Hz}$ (Model~iii); beyond these ranges, only the ABH-generated SGWB is detectable. In the lensed scenario, these ranges extend to $5{-}155\,\mathrm{Hz}$ (Model~i), $5{-}161\,\mathrm{Hz}$ (Model~ii), and $5{-}164\,\mathrm{Hz}$ (Model~iii), with only the ABH-generated SGWB detectable above the upper bounds.

In conclusion, both LIGO and ET can distinguish ABHs from PBHs via the SGWB. In the unlensed case, LIGO cannot detect the PBH-generated SGWB, resulting in a fixed distinguishable frequency range. In the lensed scenario, LIGO can detect the PBH-generated SGWB within a narrow band, and the frequency range remains nearly consistent across all three models. Since the ET-D PI curve lies below that of LIGO, two discrimination criteria apply across different frequency ranges: at lower frequencies, ABHs and PBHs can be distinguished by their SGWB strengths; at higher frequencies, any SGWB detected by ET must originate from ABHs. The frequency boundaries for these criteria are similar across the three PBH models, and the lensing effect serves to extend these boundaries.

\section{Acknowledgement}

We are grateful to Zu-Cheng Chen, Li-Ming Zheng, Huan Zhou and Dandan Lian for their helpful discussion. This work is supported by the National Key Research and Development Program of China Grant Nos. 2023YFC2206702, 2023YFC2206703 and 2021YFC2203001; National Natural Science Foundation of China under Grants Nos. 12322301, 12275021, 12073005 and 12021003; the Guangdong Basic and Applied Basic Research Foundation(Grant No. 2023A1515030116); and the Interdiscipline Research Funds of Beijing Normal University.
\bibliography{references}
\end{document}